\documentclass{article}

  \usepackage[preprint]{neurips_2025}

\usepackage[utf8]{inputenc} 
\usepackage[T1]{fontenc}    
\usepackage{hyperref}       
\usepackage{url}            
\usepackage{booktabs}       
\usepackage{amsfonts}       
\usepackage{nicefrac}       
\usepackage{microtype}      
\usepackage{xcolor}         
\usepackage{lipsum}
\usepackage{graphicx}
\usepackage{multicol}
\usepackage{float}

\title{Modelling Emotions is an Elusive Pursuit \\ in Affective Computing}

\author{%
  Anders Rolighed Larsen \\
    Department of Applied Mathematics and Computer Science\\
 Technical University of Denmark \\
 \And
Sneha Das \\
  Department of Applied Mathematics and Computer Science\\
 Technical University of Denmark \\
\texttt{sned@dtu.dk} 
\And 
 Nicole Nadine Lønfeldt \\
  Child and Adolescent Mental Health Center, Copenhagen University Hospital, Denmark \\
\texttt{nicole.nadine.loenfeldt@regionh.dk} 
\And 
 Paula Petcu \\
  Interhuman AI \\ 
  \texttt{paula@interhuman.ai}
\AND
Line Clemmensen\\
  Department of Applied Mathematics and Computer Science\\
 Technical University of Denmark\\
 \texttt{lkhc@dtu.dk} \\
}

\begin{document}

\maketitle

\begin{abstract}
  Affective computing - combining sensor technology, machine learning, and psychology - have been studied for over three decades and is employed in AI-powered technologies to enhance emotional awareness in AI systems, and detect symptoms of mental health disorders such as anxiety and depression. However, the uncertainty in such systems remains high, and the application areas are limited by categorical definitions of emotions and emotional concepts.  This paper argues that categorical emotion labels obscure emotional nuance in affective computing, and therefore continuous dimensional definitions are needed to advance the field, increase application usefulness, and lower uncertainties. 
\end{abstract}

\section{Introduction}

Affective computing systems include uni-modal research fields like facial emotion recognition (FER) \citep{FERBook}, speech emotion recognition (SER) \citep{ma2023emotion2vecselfsupervisedpretrainingspeech}, as well as multi-modal sentiment analysis (MSA) \citep{hu2024recenttrendsmultimodalaffective, SOLEYMANI20173}, and human-computer interaction \citep{SOLEYMANI20173,Preeti2012MULTIMODALER}. The emotion recognition task entails mapping input signals such as speech, facial expressions, and text to affective states. This mapping is most commonly performed using categorical emotion annotations (CEA), such as \textit{happy}, \textit{sad}, or \textit{angry}, derived from psychological taxonomies that include Ekman's basic emotions \citep{ekman1992argument} and Plutchik’s psychoevolutionary model \citep{PLUTCHIK19803} \citep{FERBook, MultimodalEmotionRecognitionusingDeepLearning_2021, SystematicAdressing}.These taxonomies serve as the primary annotation schema in many widely-used datasets, including MELD \citep{MELD}, eNTERFACE \citep{eNTERFACE}, SAVEE \citep{SAVEE}, TESS \citep{TESS}, EmoDB \citep{EmoDB}, and FER2013 \citep{FER2013}, where emotion labels are typically drawn from a fixed set of basic categories. While this approach offers simplicity of annotation and computational clarity, it imposes rigid boundaries on phenomena that are often ambiguous, overlapping, temporally dynamic and rarely experienced in isolation \citep{5349500,10.1145/3129340,
EmotionBias2, Busso2008, Bradley1999AffectiveNF, Park_2020, SOLEYMANI20173}.

The IEMOCAP dataset \citep{Busso2008}, widely used in affective computing research, exemplifies this issue. Here, each utterance is labeled with a categorical emotion based on majority voting among three annotators, while also independently annotated with continuous valence, arousal, and dominance (VAD) scores. However, the dual annotation structure reveals a critical disjunction: VAD values often deviate significantly from the emotional categories with which they are associated, and annotators frequently disagree (see Section 3 of this paper). This inconsistency, along with a lack of a general consensus definition of emotions \citep{CABANAC200269}, challenges the notion of a singular ground truth based on single categorical labels. Instead, some suggest that emotions should be modeled as distributions over possible states rather than as fixed points in a discrete label space \citep{10.1007/978-3-642-23163-6_27}.

Although datasets like IEMOCAP provide categorical labels and continuous VAD scores, these are typically used independently. The categorical label is derived via a majority vote, thereby collapsing the annotator disagreement into a singular outcome. Recent modeling strategies aim to preserve this ambiguity through soft-labeling techniques, fuzzy classifiers, or emotional profiling frameworks that treat emotional states as distributions rather than fixed categories \citep{10208626,BussoNaturalisticDataset,10.1162/tacl_a_00449}. These methods reflect a growing recognition that emotions are computationally and psychologically context dependent, temporally fluid, and rarely experienced as singular static states \citep{russell1980circumplex,EmotionBias2}.  

These challenges have prompted a growing call for alternative representations that better reflect the fluid and context-dependent nature of human emotion. Rather than treating ambiguity and disagreement as annotation noise, emerging work suggests that they are essential signals that emotion-aware systems should accommodate. 

\textbf{This paper takes the position that categorical emotion labels, while computationally convenient, obscure emotional nuance, ambiguity, and subjectivity, thus limiting the fidelity, interpretability, and ethical deployment of affective computing systems.}

\section{Theoretical Foundations and Conceptual Incongruence in Emotion Modeling}

Emotion modeling lies at the intersection of psychological theory and computational pragmatism. Although this confluence offers rich interpretative power, it also reveals fundamental tensions in how affect is conceptualized, measured, and operationalized, especially within the context of MSA. By critically examining the structure of emotional theories in psychology and the representation strategies in MSA, this section aims to expose an underlying discord: Current systems largely neglect the granularity and ambiguity inherent in real emotional experiences.

\paragraph{Contrasting Emotional Frameworks} Psychological models of emotion have historically evolved along two primary axes: discrete and dimensional. Discrete models, exemplified by Plutchik’s Wheel of Emotions \citep{PLUTCHIK19803}, Parrott’s hierarchical taxonomy \citep{Parrott2001EmotionsIS,10.1007/978-3-642-23163-6_27}, and Ekman’s universal emotions theory \citep{ekman1992argument}, categorize emotions into bounded classes. These models offer practical advantages in clarity and universality, but critics have noted their limitations in capturing the complex, often culturally mediated, nuances of emotional expression \citep{EmotionBias1, EmotionBias2}.

In contrast, dimensional models conceptualize emotions as existing on continuous spectra. The circumplex model of affect \citep{russell1980circumplex} defines emotions across the valence and arousal axes, often extended with a third dimension, such as dominance or potency \citep{Schlosberg1954ThreeDO,VADScores}. This structure supports more fluid interpretations of affect, especially useful in therapeutic or introspective settings where emotions rarely conform to single-label categories \citep{Busso2008,Bradley1999AffectiveNF, das2022continuous, das2022zero}. Further, research has also shown the language dependency of frameworks and specific factors in the frameworks \citep{9746450, hjuler2025exploring, hjuler2025emotale}. 

\paragraph{Representation Constraints in MSA} Despite the theoretical richness of the dimensional models, MSA systems have predominantly favored discrete approaches. Labels derived from Ekman's categories are easier to annotate and align with multimodal data sets \citep{8636432}, making them attractive for large-scale computational tasks. A gradual shift toward dimensional representations, typically valence, arousal, and dominance, indicates a growing awareness of the need for nuance, particularly in modeling emotional intensity or co-occurring states.

However, the scope of representation remains restricted. MSA tends to focus on prototypical emotions, happiness, sadness, anger, fear, and disgust, which produce high agreement between annotators and clearer detection signals \citep{Park_2020}. Emotions like guilt, nostalgia, or ambivalence, which reflect blended or ambiguous affective states, are frequently omitted due to annotation challenges and data sparsity. Practitioners \citep{PeterNeville} occasionally introduce custom axes such as 'Energy' or 'Tension' to refine the granularity, although such choices often stem from subjective preference rather than theoretical consistency.

This pragmatic divergence leads to a critical distinction: psychological inquiry prioritizes subjective experience and emotional complexity, while MSA is designed around observable signals and model efficiency. Consequently, the emotional spectrum in MSA is operationalized in a narrower, often oversimplified format that hampers the recognition of subtle or complex emotions \citep{SOLEYMANI20173}.

\paragraph{Conceptual Vagueness and Terminological Diffusion} The representational gap between psychological theory and computational practice is further compounded by the lack of terminological clarity in affective computing. 
Affective computing lacks a shared vocabulary for describing emotional ambiguity. In the literature, a wide range of loosely defined terms, as shown in Figure~\ref{fig:emotion_ambiguity_terms}, are used to describe similar phenomena of interpretive uncertainty without clear agreement. Some of these terms originate from psychological theory, others from system design, but their inconsistent use has prevented convergence toward a common taxonomy.


Of particular relevance is the notion of \textit{emotional ambivalence}, defined as the co-occurrence of opposing emotional states toward the same stimulus \citep{Larsen2001CanPF,NeuroLaunch,10.1093/acprof:oso/9780190613501.003.0018}. It also mirrors the annotation behavior seen in evaluators who assign multiple labels to a single utterance, revealing uncertainty or a nuanced perception rather than indecision. Even the person experiencing ambivalence may have trouble describing the experience.

Adjacent to this is the concept of \textit{prototypicality}, where consensus in annotation signals a normative emotional expression. Conversely, \textit{non-prototypicality} and \textit{emotional incongruity} challenge standard classification models by surfacing the ambiguity in cross-modal cues, for example, when the vocal tone contradicts facial expression \citep{10.1145/2993148.2993173,7344624,10.1007/978-3-642-23163-6_27,Deng_2021_ICCV}. Annotators may differ in their ability to express own or read others' emotions, in which case striving for a consensus may be seen as averaging out noise. Or it could be seen as a more continuous and natural variation in perception, in which case using distributions to model emotions seems like the better option. Finally, ambiguity across modalities may reflect more complex signals, like sarcasm, suppressed emotions, or situations where the person in question has multiple emotions or transitions between emotions.

In light of this terminological diffusion, we adopt the term \textit{emotional ambiguity} to unify the various phenomena described. Unlike terms that focus solely on polarity or disagreement, emotional ambiguity captures both the presence of overlapping or conflicting emotional cues and the interpretive uncertainty they provoke. It emphasizes the inherently fluid, multi-layered nature of affective perception—particularly when signals across modalities or annotators diverge.

\paragraph{Ambiguity in Emotion Datasets} Despite increasing interest in emotional nuance, the representation of ambiguity in publicly available datasets remains inconsistent. Standard corpora such as IEMOCAP \citep{Busso2008} offer high-quality multimodal data, but largely emphasize dominant emotion labels, often masking the presence of mixed or uncertain affective cues. Although some follow-up work notes the existence of mixed emotions utterances \citep{5349500}, the annotation protocol itself discourages capturing the complexity found in spontaneous expression. Tran et al. \citep{EnglishAmbiguity,FrenchAmbiguity}observe that the expressions acted in IEMOCAP tend to overrepresent prototypical emotions, thus underrepresenting the interpretive ambiguity characteristic of naturalistic interaction.

Data sets such as CMU-MOSEI \citep{bagher-zadeh-etal-2018-multimodal} and CMU-MOSEAS \citep{bagher-zadeh-etal-2020-cmu} offer finer granularity by allowing per-emotion intensity ratings. However, even in these cases, the inherent ambiguity in the data points is often overlooked or relegated to future work considerations \citep{EnglishAmbiguity,FrenchAmbiguity,AmbiguitySurveyPriya,PAN2023126866,s23115184,mti6060047,ambiguitySurveyRai,shou2023comprehensivesurveymultimodalconversational}. This reveals a broader trend: while datasets may technically permit multi-label or distributional annotation, prevailing usage patterns default to hard-label classification that suppresses emotional variability.

\paragraph{Annotation Practices and Distributional Alternatives} These limitations are not only a matter of dataset design but also of annotation strategy. Majority voting schemes remain standard, despite evidence that such aggregation often conceals meaningful inter-annotator disagreement \citep{davani-etal-2022-dealing,fleisig-etal-2023-majority}. Such disagreements are not merely noise; they reflect socio-cultural diversity, perceptual idiosyncrasies, and inherent emotional ambiguity \citep{SocialCulturalDifferences,Scherer1994,plisiecki2024highriskpoliticalbias,8682170}. Systems that treat these disagreements as error signals risk misrepresenting emotion as an objective truth rather than a subjective construct.

Recent work offers promising alternatives. Soft-label distributions, Likert scales \citep{10208626,LikertScale}, and fuzzy emotion models \citep{10.1007/978-3-642-23163-6_27} allow expressions to exist in a multidimensional affective space. Emotional profiling, proposed by Mower et al. \citep{5349500}, represents emotional states as distributions between categories. Complementing this, emotional interpolation techniques weigh utterances in relation to the broader context of a dialogue, allowing for ambiguity without forcing artificial resolution.

Crowdsourcing methodologies have advanced similarly, particularly in how they accommodate emotional ambiguity. Mariooryad et al. \citep{BussoNaturalisticDataset} proposed a three-stage Likert-based AMT pipeline designed to generate more nuanced emotional judgments by capturing individual perspectives of annotators rather than collapsing them into consensus. Building on this notion of preserving subjectivity, recent multitask learning approaches have introduced models that explicitly account for annotator-specific bias by learning separate label distributions for each rater alongside aggregate consensus labels \citep{davani-etal-2022-dealing,8682170,snow-etal-2008-cheap}. These architectures not only maintain the diversity of emotional interpretation, but also demonstrate improved recall and greater interpretability, particularly in the presence of ambiguity.

Debate is emerging around the legitimacy of human annotation altogether. Plisiecki et al.\citep{plisiecki2024highriskpoliticalbias} warn that political and ideological biases embedded in annotators can corrupt model output in downstream tasks, advocating for lexicon-driven alternatives. Self-annotation, as used in IEMOCAP and later revisited by Zhang et al.\citep{
10.1145/2993148.2993173} and Saganowski et al. \citep{Saganowski2022EmognitionDE}, offers one way to address this, capturing the introspective dimensions of affective experience - although this is not without its own set of limitations.

\paragraph{Pragmatic Merits of Categorical Annotation} 
Despite their limitations, categorical emotion annotations remain widely used in affective computing due to their operational efficiency and empirical robustness in constrained tasks. Discrete categories such as \textit{happy}, \textit{angry}, and \textit{sad} offer a shared vocabulary between annotators and systems, simplifying the annotation and model evaluation. As \citep{10.1145/3129340} highlights in a retrospective review of speech emotion recognition, the field has made significant strides using categorical labels as benchmarks, allowing standardized comparisons and cumulative progress in studies. In emotion classification tasks, where rapid system response or interpretability is essential, such as driver monitoring \citep{app12020807, EspinoSalinas2024} or call center analytics \citep{DBLP:journals/corr/abs-2110-14957}, categorical predictions serve actionable outputs that are easy to map to rules or interventions. Moreover, categorical models often outperform dimensional ones in low-data regimes or when emotion intensity is low, as shown in \citep{10.1145/2993148.2993173}'s joint modeling of perceived and self-reported emotion. Categorical annotation schemes also facilitate multimodal data fusion: since modalities may align more naturally to prototypical categories than abstract affective spaces, joint modeling becomes more tractable. Even when emotional states are complex, annotators can provide multiple labels or rank primary versus secondary emotions, which is a practical compromise seen in datasets such as IEMOCAP \citep{Busso2008}.

\subsection{Position and Outlook}

The empirical and theoretical evidence presented throughout this paper points to a central problem: affective computing systems that rely on categorical emotion labels are structurally limited in their ability to capture the ambiguity and subjectivity inherent in emotional expression.
This limitation is a fundamental mismatch between the complexity of affective experience and the representational tools we use to capture it. Rather than treating ambiguity as noise or error, we argue that it should be embraced as a modeling target. In the sections that follow, we explore the consequences of this misalignment for real-world systems, including risks to interpretability, ethical use, and user trust.

\section{Empirical Breakdown of Categorical Emotion Assumptions}

To ground the theoretical critiques outlined above, we present empirical analyses of the IEMOCAP dataset \citep{Busso2008}, which provides both CEA and continuous VAD scores. These analyses expose multiple forms of annotation misalignment that call into question the reliability and expressiveness of discrete labels.

\begin{figure}[!ht]
    \centering
    \includegraphics[width=0.7\linewidth]{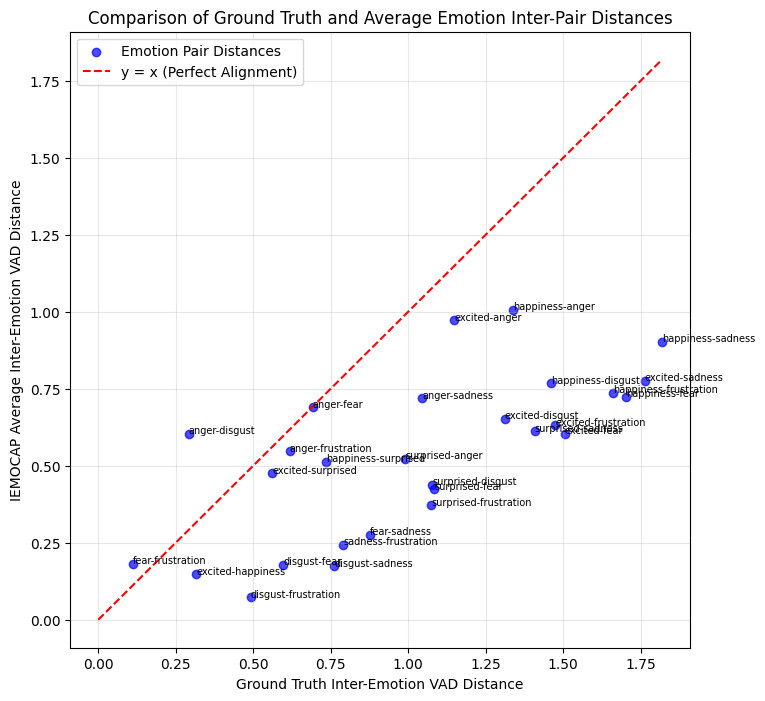}
   \caption{Pairwise emotion distances in VAD space: theoretical ground truth vs. IEMOCAP annotator averages. Most points fall below the diagonal, indicating that annotators perceived emotion categories as more similar than dimensional theory predicts.}
    \label{fig:vad-theory-empirical}
\end{figure}

\paragraph{Dimensional Inconsistencies within Categorical Labels.}  
Although each IEMOCAP utterance is assigned a single categorical label (e.g., \textit{sad}, \textit{angry}), their associated VAD values show considerable dispersion even within the same label class. When these empirical VAD annotations are compared with theoretically expected VAD coordinates (as derived from Russell's three-factor model), we observe significant divergence patterns. For example, utterances labeled \textit{ happy} span a wide range of arousal and dominance, violating the premise of internal VAD coherence within a category. Figure~\ref{fig:vad-theory-empirical} visualizes this divergence, plotting empirical and theoretical VAD positions, and highlighting how averaged annotator VAD scores deviate from canonical emotion coordinates. This discrepancy suggests that categorical labels often conceal heterogeneous emotional interpretations.

\begin{figure}[!ht]
    \centering
    \includegraphics[width=1.0\linewidth]{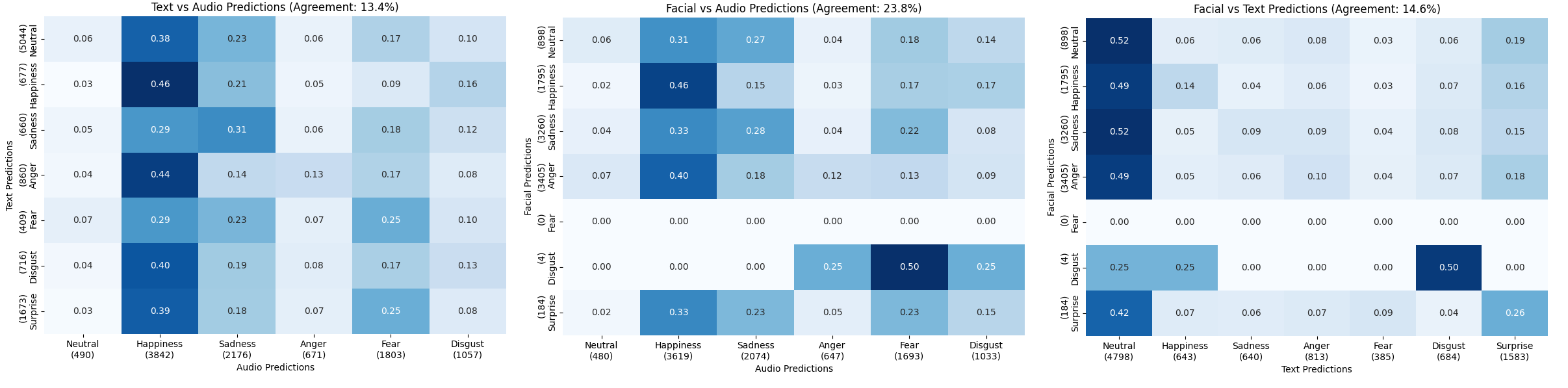}
   \caption{
    Cross-modal prediction alignment matrices for each modality pair, with overall agreement percentages indicated in parentheses. Left: Text vs.\ Audio predictions. Center: Facial vs.\ Audio predictions. Right: Facial vs.\ Text predictions. Each matrix is normalized by the number of predictions per class, by the y-axis predictor, to account for imbalanced emotion distributions across modalities. A difference in class support can be observed between modality comparisons, due to faulty data-entries to the facial modality. 
    }    
    \label{fig:modality-agreement}
\end{figure}

\paragraph{Modality-Specific Disagreement.}  
A second axis of ambiguity emerges when comparing emotion predictions across modalities. We report a systematic disagreement between text, audio, and visual models tested on the same utterances.Confusion matrices show, for example, that the text modality frequently defaults to \textit{neutral}, while the audio model tends to prefer \textit{sadness} or \textit{anger}, even for the same input. The general agreement between the three modalities was only 4.18\%, with 54\% of the utterances receiving completely distinct predictions in all models. These findings are summarized in Figure~\ref{fig:modality-agreement}, emphasizing that affective perception is highly modality-dependent and that categorical alignment across channels is the exception, not the norm. Since the pre-trained audio model does not include “disgust” among its recognized categories, this class is excluded from its predictions and therefore absent in the corresponding comparison matrices. Agreement metrics are calculated based solely on samples for which both modalities yield predictions within their shared label space.

\begin{table}[!ht]
\centering
\caption{Overview of pre-trained emotion recognition models used for each modality.}
\label{tab:modality-models}
\begin{tabular}{@{}lp{5.25cm}p{6.5cm}@{}}
\toprule
\textbf{Modality} & \textbf{Architecture} & \textbf{Model} \\
\midrule
Text   & Transformer (DistilRoBERTa) & \texttt{emotion-english-distilroberta-base} \cite{hartmann2022emotionenglish} \\
Audio  & Transformer (Wav2Vec2)       & \texttt{w2v-speech-emotion-recognition} \cite{khoa2024emotionenglishAudio} \\
Facial & CNN + LSTM (ResNet50 + LSTM) & \texttt{EMO-AffectNetModel} \cite{RYUMINA2022} \\
\bottomrule
\end{tabular}
\end{table}

\begin{figure}[!ht]
    \centering
    \includegraphics[width=0.95\linewidth]{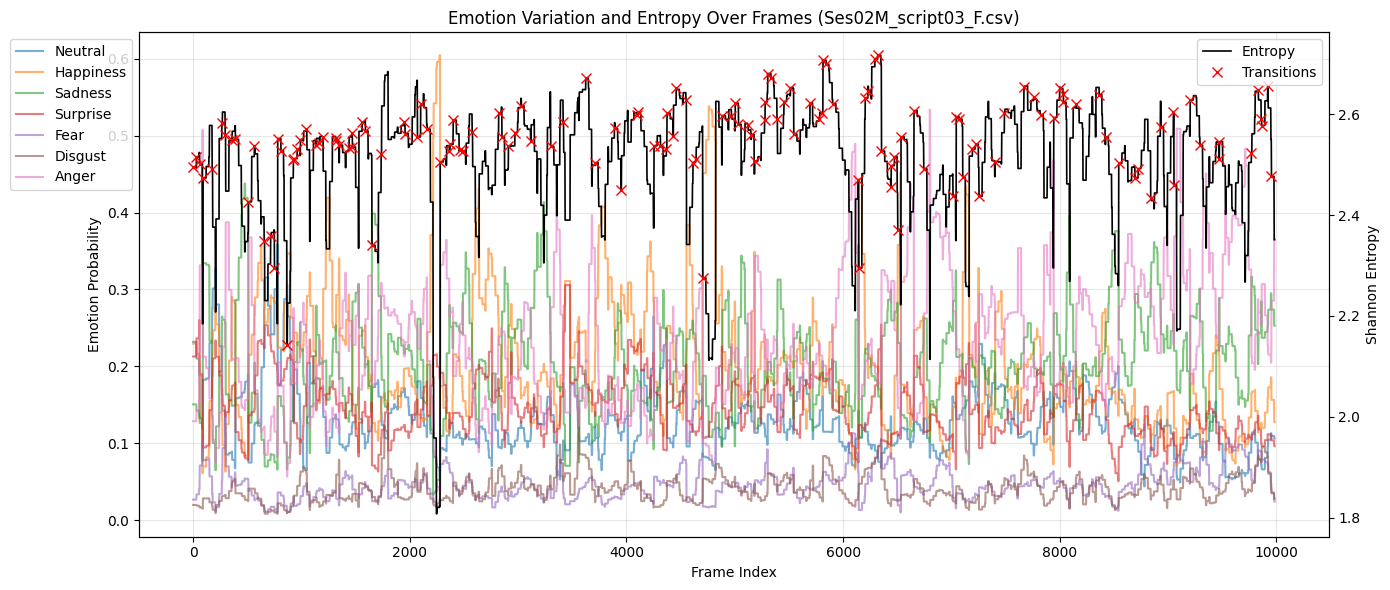}
    \caption{Framewise emotion probabilities (colored lines) and corresponding entropy (black line) for a representative utterance. Red x's mark emotion transitions. Note the alignment of entropy spikes with transitions, indicating periods of increased affective ambiguity.}
    \label{fig:frame-dynamics}
\end{figure}

\paragraph{Temporal Flattening of Emotional Transitions.}
Emotional states are not static; they often evolve substantially within the span of a single utterance. Frame-level analyses of facial expressions reveal that predicted emotional states frequently shift mid-utterance. Figure~\ref{fig:frame-dynamics} illustrates this dynamic: emotion probability curves fluctuate over time, while shannon entropy on model probabilities (black line) captures moment-to-moment uncertainty. Red markers indicate identified transition points between predicted dominant emotions, which tend to align with localized spikes in entropy. This suggests that transitions correspond with heightened affective ambiguity.

\begin{figure}[!ht]
    \centering
    \includegraphics[width=0.95\linewidth]{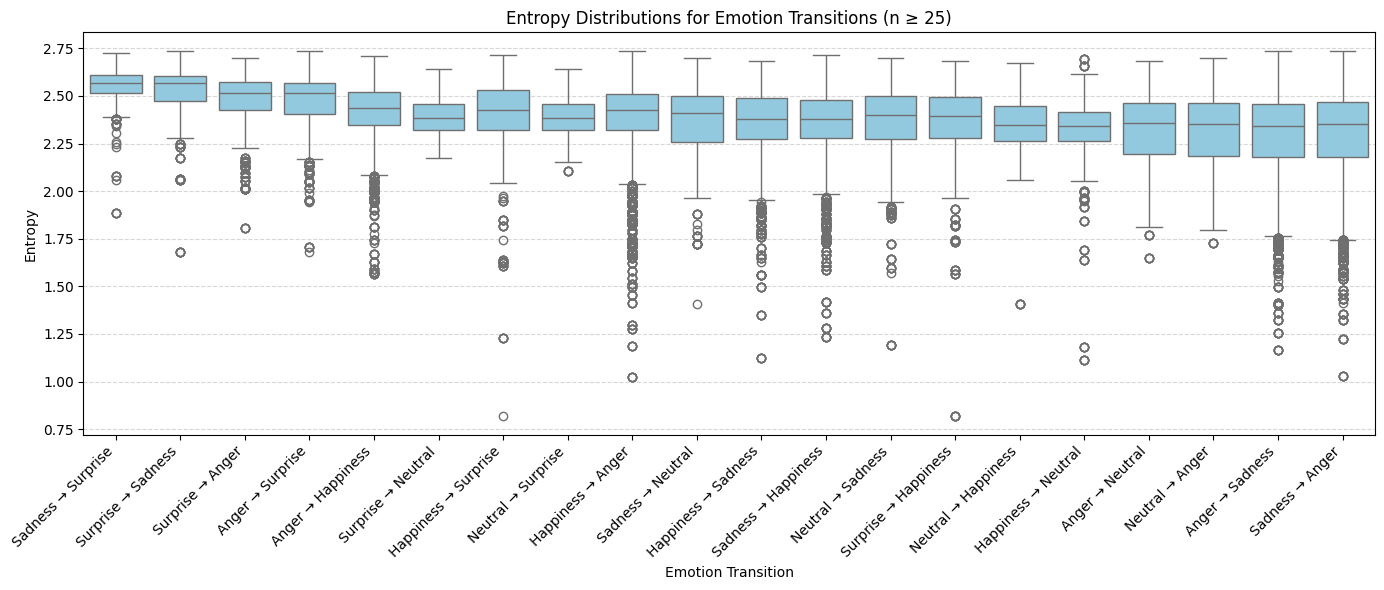}
    \caption{
    Entropy distributions across specific emotion-to-emotion transitions, based on frame-level predictions. 
    Each boxplot represents a directional transition (e.g., \textit{sadness}~$\rightarrow$~\textit{anger}) with at least 25 occurrences. Transitions involving \textit{surprise}, \textit{neutral}, or cross-valence shifts tend to show elevated entropy, suggesting some emotion shifts are inherently more ambiguous than others.
    }
    \label{fig:entropy_transitions}
\end{figure}

This local pattern holds at scale. As shown in Appendix Figure~\ref{fig:entropy_global}, entropy across all transitional frames ($\pm$ 5 frames at transition point) is significantly higher than during baseline (stable sequences) periods, with a strong statistical effect (p~$<<$~0). There was discarded 5 frames between a transition window and a stable sequence in an attempt to alleviate auto-correlation. This global analysis confirms that emotion transitions are not just points of label change, but regions of elevated model uncertainty, stemming from the constraints placed by the categorical annotation domain. Moreover, entropy varies across specific emotion shifts: transitions involving surprise, neutrality, or shifts between positively and negatively valenced states exhibit the highest entropy (Appendix Figure~\ref{fig:entropy_transitions}). These findings underscore that not all transitions are equally ambiguous, and that ambiguity itself is a function of emotional context and trajectory.

Taken together, these results demonstrate that flattening entire utterances into single categorical labels discards rich affective dynamics and obscures the interpretive ambiguity inherent in transition phases. Emotion-aware systems would benefit from modeling such dynamics explicitly, whether via frame-level representations, temporal smoothing, or probabilistic tracking of multiple concurrent emotional states.

\section{Evaluation Under Ambiguity}

Emotion recognition systems are typically benchmarked against discrete ground truth labels, assuming these annotations are accurate and representative of a singular emotional state. However, the findings of the IEMOCAP dataset challenge this assumption. Full annotator agreement on categorical emotion annotations (CEA) was observed in only approximately 20\% of utterances, and nearly 25\% of utterances lacked a clear majority label overall. These rates indicate a substantial degree of inherent ambiguity in the data itself.

\subsection{Impact of Ambiguity on Model Performance}

To quantify how ambiguity influences model evaluation, we implemented a filtering strategy based on categorical agreement levels. This method creates partitions of the data set based on the proportion of annotators who agreed on the majority label for each utterance. For example, a threshold of 0.75 retains only those samples in which at least 75\% of the annotators selected the same emotion, thereby reducing the influence of ambiguous or contested labels.Model performance was evaluated using the weighted F1 score, which averages per-class F1 scores weighted by label frequency to account for class imbalance. As ambiguity decreased (i.e., with more stringent thresholds), model performance consistently improved for audio and facial modalities. For example, the weighted F1 of the audio model increased from 0.332 at baseline to 0.496 at higher level of prototypicality, indicating a substantial sensitivity to the clarity of the annotation. The facial model also showed marked improvements. In contrast, the performance of the text modality remained relatively stable, highlighting its greater resilience to label noise, but also possibly its lower sensitivity to affective nuance.

\begin{figure}[!ht]
    \centering
    \includegraphics[width=0.75\linewidth]{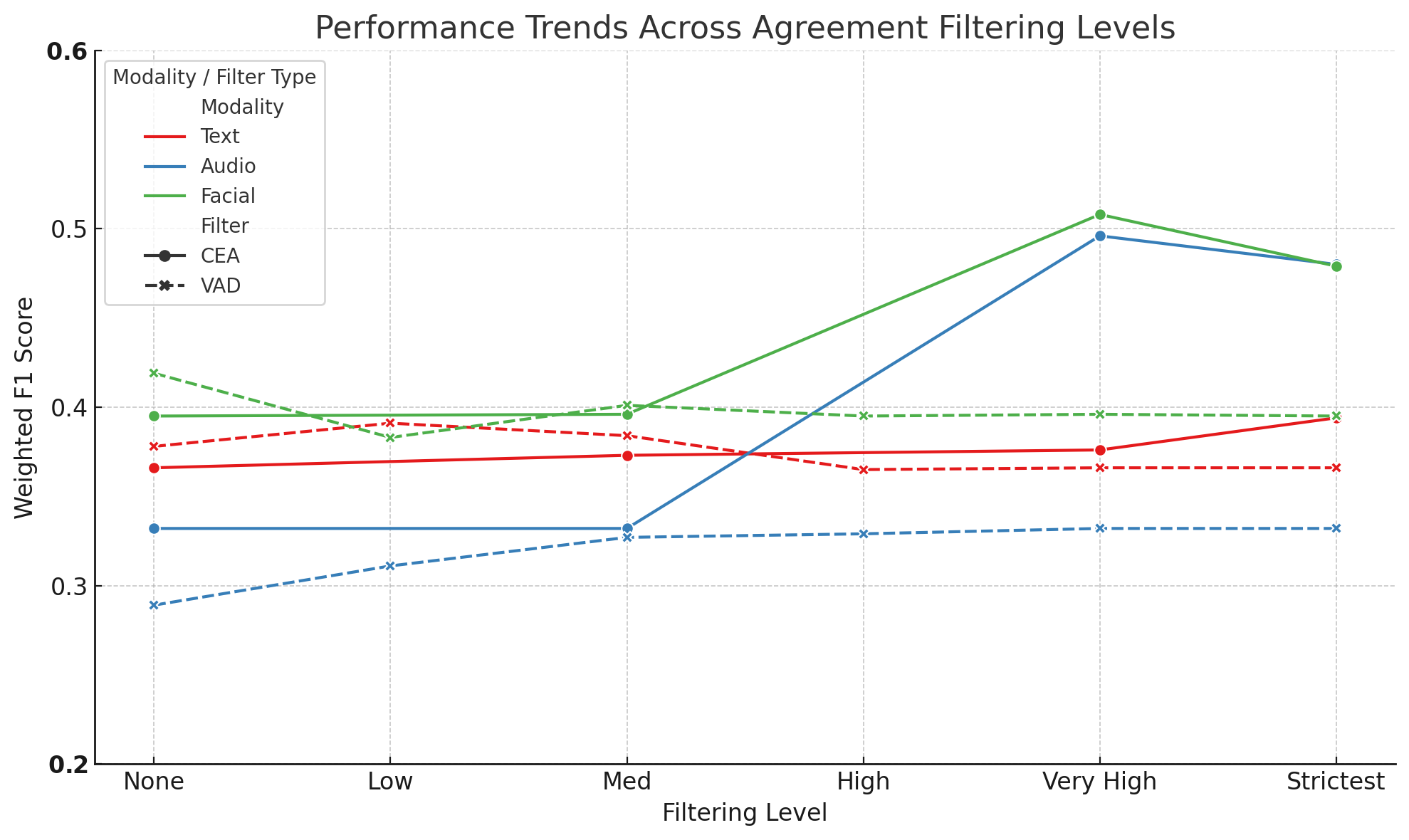}
    \caption{Weighted F1 scores for text, audio, and facial emotion recognition models across increasing levels of agreement-based filtering. 
    Filtering is applied based on either categorical emotion annotation (CEA, solid lines) or VAD score coherence (VAD, dashed lines). Values and partioning parameters can be found in Appendix Figures~\ref{tab:cea_f1_scores} and \ref{tab:vad_f1_scores}.}
    \label{fig:f1-thresholding}
\end{figure}

These effects are illustrated in Figure~\ref{fig:f1-thresholding}, which shows weighted F1 scores at increasing levels of cut-off values used for CEA agreement. The trend suggests that high-agreement data provide more consistent signals and a cleaner evaluation, with the implication that data, when ambiguous, is also filtered out in the process.

\subsection{VAD Dispersion as an Alternative Metric}

An alternative ambiguity filter was tested using the dispersion in the euclidean distance space of the valence, arousal, and dominance scores, in the [0,5] domain as is the range used in IEMOCAP. Contrary to expectations, reducing VAD disagreement did not consistently improve model performance. For the audio modality, the performance decreased slightly with stricter VAD agreement filters, and the performance of the facial modality remained flat or inconsistent. This finding suggests that continuous VAD annotations from the IEMOCAP dataset, despite their finer granularity, do not necessarily correlate with categorical ground-truth reliability or clarity. In addition, it reinforces the misalignment between the two annotation paradigms.

\subsection{Agreement Across Modalities}

Model disagreement further complicates the evaluation landscape. Full agreement among the three unimodal models—text, audio, and facial—was achieved in just 4.18\% of the utterances. This low concordance persisted even when the predictions were compared pair-wise, with a particularly weak alignment between the text and audio modalities. The divergence illustrates that each modality captures distinct affective signals and that rigid alignment to a singular label may misrepresent this variation.

Table~\ref{tab:modality-agreement} summarizes the frequency of complete agreement between the models and highlights the emotions that are the most agreed on.

\begin{table}[!ht]
    \centering
    \caption{Emotion-wise agreement count across modalities (text, audio, facial).}
    \label{tab:modality-agreement}
    \begin{tabular}{lccc}
        \toprule
        Emotion & Agreed Count & Share of Total \\
        \midrule
        Happiness & 125 & \\
        Sadness & 104 & \\
        Anger & 67 & \\
        Neutral & 37 & \\
        \textbf{Total (All Emotions)} & \textbf{333} & \textbf{4.18\%} \\
        \bottomrule
    \end{tabular}
\end{table}

This systemic misalignment, between models, annotations, and modalities, undermines the validity of single-label evaluations and calls for a rethinking of how emotion recognition systems are trained and assessed. Categorical evaluation overlooks perceptual divergence, while VAD ratings, though more granular, fail to provide a reliable proxy for consensus. Together, these findings suggest that ambiguity is not only noise to be removed, but also contains a signal to be modeled.

\section{Ambiguity in Practice: Ethical and Experimential Implications}

Emotion recognition technologies are increasingly being deployed in settings where outputs are not merely academic predictions, but guide real-time feedback, decision-making, and interpersonal understanding. Applications such as coaching, education, and mental health support place these systems in ethically sensitive roles, where labeling affect carries practical and psychological consequences. However, most current systems, often trained on majority-voted categorical labels, are structurally predisposed to flatten complexity.

A key concern is that such systems convey a false sense of emotional precision, masking underlying uncertainty or disagreement. Systems typically output singular emotion labels such as \textit{happy} or \textit{sad} without communicating underlying ambiguity, disagreement, or temporal fluctuation. Figure~\ref{fig:modality-agree-fails} illustrates one such case: all three unimodal models converged on the same label (\textit{happiness}), despite annotators labeling the utterance as \textit{anger} or \textit{frustration}. This divergence, often tied to sarcasm, ambivalence, or multimodal mismatch, reveals how overconfident outputs can misrepresent the user’s state.

\begin{figure}[!ht]
    \centering
    \includegraphics[width=1.0\linewidth]{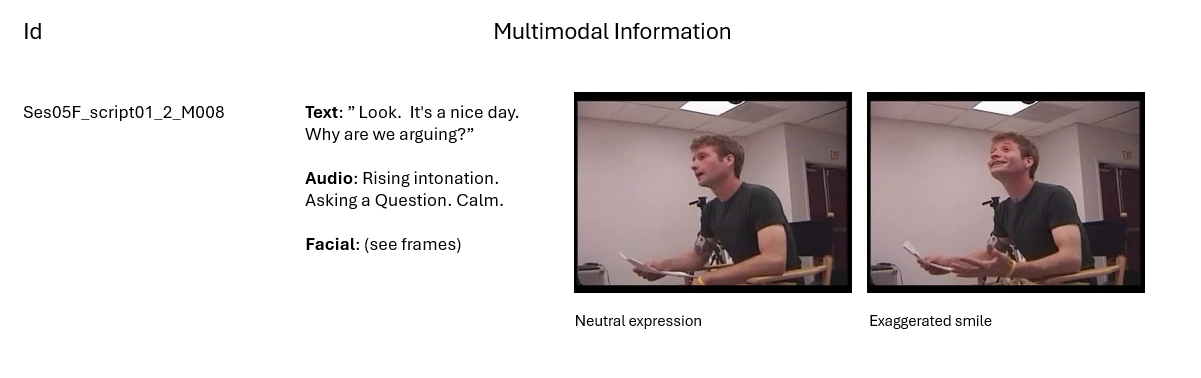}
    \caption{Case of modality agreement in system predictions diverging from ground truth. All models predicted \textit{happy}, while annotators labeled the utterance as \textit{angry} or \textit{frustrated}. Examples directly taken from IEMOCAP data set \citep{Busso2008} and used under their licensing agreement.}
    \label{fig:modality-agree-fails}
\end{figure}

Such misclassifications are not neutral. In emotionally sensitive domains, users may feel unseen, invalidated, or misunderstood by systems that present misleadingly precise feedback. Over time, these mismatches can erode user trust and reduce the perceived empathy or usefulness of the technology - especially in longitudinal or therapeutic settings. When ambiguity is suppressed through majority voting or softmax outputs, interpretive nuance is lost.

Systems deployed in emotionally responsive settings must not only detect affect but track its evolution over time. Failing to register how emotional states build, shift, or resolve risks producing responses that feel uncalibrated or disconnected from lived experience. This is especially problematic in dialogue-based applications, where a flattened emotional summary can obscure meaningful turning points. A system that fails to register rising frustration, for instance, may miss the opportunity for timely intervention. As earlier analyses showed (Figure~\ref{fig:frame-dynamics}), these dynamics are perceptible at the frame level, yet are often lost in current labeling approaches.

Ambiguity in emotion data is not a defect to be corrected but a core feature of human expression. Systems that erase this complexity in pursuit of clarity risk technical inaccuracy and experiential harm. Ethically, developers of emotion-aware AI must treat labeling as an interpretive act — not merely a prediction task. Practically, users deserve systems that reflect the nuance and fluctuation inherent in how emotions are expressed and perceived.

\section{Position and future directions}

Emotion recognition systems today continue to rely on categorical annotations that reduce complex, context-dependent affective states to singular labels. As demonstrated throughout this paper, this simplification overlooks valuable information, such as variability between annotators, divergence between modalities, and transitions within utterances, that could otherwise inform more nuanced and robust models.

We do not claim that dimensional or fuzzy approaches are fully mature alternatives. However, we argue that the field must begin treating ambiguity as an informative signal rather than noise to be discarded. Moving forward, promising directions include the adoption of soft-label distributions, emotional profiling, and annotator-specific modeling, all of which preserve subjectivity rather than collapsing it. Similarly, integrating contextual features such as dialogue structure, temporal dynamics, and multiple perceptual modalities will allow systems to infer emotional states more fluidly and adaptively—capturing the layered cues that contribute to ambiguity in human emotional perception.

The shift ahead is not merely technical, but rather it is conceptual. Affective computing must revise its assumptions about emotion itself: from discrete to continuous, from static to situated, from certain to interpretively plural. Embracing this complexity, and exploring new concepts, is the necessary next step if emotion-aware systems are to reflect the nuance and variability of human affective experience.

\bibliographystyle{plainnat}  
\bibliography{bibliography}      

@article{CABANAC200269,
title = {What is emotion?},
journal = {Behavioural Processes},
volume = {60},
number = {2},
pages = {69-83},
year = {2002},
issn = {0376-6357},
doi = {https://doi.org/10.1016/S0376-6357(02)00078-5},
url = {https://www.sciencedirect.com/science/article/pii/S0376635702000785},
author = {Michel Cabanac}
}

@article{ekman1992argument,
  author = {Ekman, Paul},
  journal = {Cognition \& emotion},
  number = {3-4},
  pages = {169--200},
  publisher = {Taylor \& Francis},
  title = {An argument for basic emotions},
  volume = 6,
  year = 1992
}

@article{russell1980circumplex,
  author = {Russell, J.A.},
  issn = {0022-3514},
  journal = {Journal of personality and social psychology},
  number = 6,
  pages = {1161--1178},
  title = {{A circumplex model of affect}},
  volume = 39,
  year = 1980
}

@article{EmotionBias1,
author = {Cordaro, Daniel and Sun, Rui and Keltner, Dacher and Kamble, Shanmukh and Huddar, Niranjan and McNeil, Galen},
year = {2017},
month = {06},
pages = {},
title = {Universals and Cultural Variations in 22 Emotional Expressions Across Five Cultures},
volume = {18},
journal = {Emotion},
doi = {10.1037/emo0000302}
}

@article{EmotionBias2,
author = {Gendron, Maria and Crivelli, Carlos and Barrett, Lisa},
year = {2018},
month = {07},
pages = {096372141774679},
title = {Universality Reconsidered: Diversity in Making Meaning of Facial Expressions},
volume = {27},
journal = {Current Directions in Psychological Science},
doi = {10.1177/0963721417746794}
}

@ARTICLE{8636432,
  author={Poria, Soujanya and Majumder, Navonil and Hazarika, Devamanyu and Cambria, Erik and Gelbukh, Alexander and Hussain, Amir},
  journal={IEEE Intelligent Systems}, 
  title={Multimodal Sentiment Analysis: Addressing Key Issues and Setting Up the Baselines}, 
  year={2018},
  volume={33},
  number={6},
  pages={17-25},
  doi={10.1109/MIS.2018.2882362}}

@article{Larsen2001CanPF,
  title={Can people feel happy and sad at the same time?},
  author={Jeff T. Larsen and A. Peter McGraw and John T. Cacioppo},
  journal={Journal of personality and social psychology},
  year={2001},
  volume={81 4},
  pages={
          684-96
        },
  url={https://api.semanticscholar.org/CorpusID:18806410}
}

@misc{NeuroLaunch,
  title = {Emotional Ambivalence},
  howpublished = {\url{https://neurolaunch.com/emotional-ambivalence/}},
  note = {Accessed: 27-01-2025}, 
  year = {2024},
  author = {NeuroLaunch}
}

@inproceedings{Bradley1999AffectiveNF,
  title={Affective Norms for English Words (ANEW): Instruction Manual and Affective Ratings},
  author={Margaret M. Bradley and Peter J. Lang},
  year={1999},
  url={https://api.semanticscholar.org/CorpusID:145474983}
}

@article{Busso2008,
  author = {Busso, Carlos and Bulut, Murtaza and Lee, Chi-Chun and Kazemzadeh, Abe and Mower, Emily and Kim, Samuel and Chang, Jeannette N. and Lee, Sungbok and Narayanan, Shrikanth S.},
  journal = {Language Resources and Evaluation},
  month = dec,
  number = 4,
  pages = {335-359},
  title = {{IEMOCAP}: interactive emotional dyadic motion capture database},
  volume = 42,
  year = 2008
}

@article{Schlosberg1954ThreeDO,
  title={Three dimensions of emotion.},
  author={Harold Schlosberg},
  journal={Psychological review},
  year={1954},
  volume={61 2},
  pages={
          81-8
        },
  url={https://api.semanticscholar.org/CorpusID:27914497}
}

@inproceedings{FrenchAmbiguity,
author = {Tran, Hélène and Brelet, Lisa and Falih, Issam and Goblet, Xavier and Mephu Nguifo, Engelbert},
year = {2022},
month = {01},
pages = {},
title = {L'ambiguïté dans la représentation des émotions : état de l'art des bases de données multimodales}
}

@inproceedings{EnglishAmbiguity,
author = {Tran, Hélène and Falih, Issam and Goblet, Xavier and Mephu Nguifo, Engelbert},
year = {2022},
month = {06},
pages = {},
title = {Do Multimodal Emotion Recognition Models Tackle Ambiguity?}
}

@inproceedings{bagher-zadeh-etal-2018-multimodal,
    title = "Multimodal Language Analysis in the Wild: {CMU}-{MOSEI} Dataset and Interpretable Dynamic Fusion Graph",
    author = "Bagher Zadeh, AmirAli  and
      Liang, Paul Pu  and
      Poria, Soujanya  and
      Cambria, Erik  and
      Morency, Louis-Philippe",
    editor = "Gurevych, Iryna  and
      Miyao, Yusuke",
    booktitle = "Proceedings of the 56th Annual Meeting of the Association for Computational Linguistics (Volume 1: Long Papers)",
    month = jul,
    year = "2018",
    address = "Melbourne, Australia",
    publisher = "Association for Computational Linguistics",
    url = "https://aclanthology.org/P18-1208/",
    doi = "10.18653/v1/P18-1208",
    pages = "2236--2246"
}

@inproceedings{bagher-zadeh-etal-2020-cmu,
    title = "{CMU}-{MOSEAS}: A Multimodal Language Dataset for {S}panish, {P}ortuguese, {G}erman and {F}rench",
    author = "Bagher Zadeh, AmirAli  and
      Cao, Yansheng  and
      Hessner, Simon  and
      Liang, Paul Pu  and
      Poria, Soujanya  and
      Morency, Louis-Philippe",
    editor = "Webber, Bonnie  and
      Cohn, Trevor  and
      He, Yulan  and
      Liu, Yang",
    booktitle = "Proceedings of the 2020 Conference on Empirical Methods in Natural Language Processing (EMNLP)",
    month = nov,
    year = "2020",
    address = "Online",
    publisher = "Association for Computational Linguistics",
    url = "https://aclanthology.org/2020.emnlp-main.141/",
    doi = "10.18653/v1/2020.emnlp-main.141",
    pages = "1801--1812"
}

@article{10.1162/tacl_a_00449,
    author = {Davani, Aida Mostafazadeh and Díaz, Mark and Prabhakaran, Vinodkumar},
    title = {Dealing with Disagreements: Looking Beyond the Majority Vote in Subjective Annotations},
    journal = {Transactions of the Association for Computational Linguistics},
    volume = {10},
    pages = {92-110},
    year = {2022},
    month = {01},
    issn = {2307-387X},
    doi = {10.1162/tacl_a_00449},
    url = {https://doi.org/10.1162/tacl\_a\_00449},
    eprint = {https://direct.mit.edu/tacl/article-pdf/doi/10.1162/tacl\_a\_00449/1986597/tacl\_a\_00449.pdf},
}

@Article{s23115184,
AUTHOR = {Aguilera, Ana and Mellado, Diego and Rojas, Felipe},
TITLE = {An Assessment of In-the-Wild Datasets for Multimodal Emotion Recognition},
JOURNAL = {Sensors},
VOLUME = {23},
YEAR = {2023},
NUMBER = {11},
ARTICLE-NUMBER = {5184},
URL = {https://www.mdpi.com/1424-8220/23/11/5184},
PubMedID = {37299912},
ISSN = {1424-8220},
DOI = {10.3390/s23115184}
}

@inproceedings{fleisig-etal-2023-majority,
    title = "When the Majority is Wrong: Modeling Annotator Disagreement for Subjective Tasks",
    author = "Fleisig, Eve  and
      Abebe, Rediet  and
      Klein, Dan",
    editor = "Bouamor, Houda  and
      Pino, Juan  and
      Bali, Kalika",
    booktitle = "Proceedings of the 2023 Conference on Empirical Methods in Natural Language Processing",
    month = dec,
    year = "2023",
    address = "Singapore",
    publisher = "Association for Computational Linguistics",
    url = "https://aclanthology.org/2023.emnlp-main.415/",
    doi = "10.18653/v1/2023.emnlp-main.415",
    pages = "6715--6726"
}

@article{SocialCulturalDifferences,
author = {Okur, Eda and Aslan, Sinem and Alyuz, Nese and Arslan, Asli and Baker, Ryan},
year = {2018},
month = {12},
pages = {},
title = {The Importance of Socio-Cultural Differences for Annotating and Detecting the Affective States of Students}
}

@article{davani-etal-2022-dealing,
    title = "Dealing with Disagreements: Looking Beyond the Majority Vote in Subjective Annotations",
    author = "Mostafazadeh Davani, Aida  and
      D{\'i}az, Mark  and
      Prabhakaran, Vinodkumar",
    editor = "Roark, Brian  and
      Nenkova, Ani",
    journal = "Transactions of the Association for Computational Linguistics",
    volume = "10",
    year = "2022",
    address = "Cambridge, MA",
    publisher = "MIT Press",
    url = "https://aclanthology.org/2022.tacl-1.6/",
    doi = "10.1162/tacl_a_00449",
    pages = "92--110"
}

@INPROCEEDINGS{10208626,
  author={Palotti, Joao and Narula, Gagan and Raheem, Lekan and Bay, Herbert},
  booktitle={2023 IEEE/CVF Conference on Computer Vision and Pattern Recognition Workshops (CVPRW)}, 
  title={Analysis of Emotion Annotation Strength Improves Generalization in Speech Emotion Recognition Models}, 
  year={2023},
  volume={},
  number={},
  pages={5829-5837},
  doi={10.1109/CVPRW59228.2023.00619}}

@article{VADScores,
author = {Russell, James and Mehrabian, Albert},
year = {1977},
month = {09},
pages = {273-294},
title = {Evidence for a Three-Factor Theory of Emotions},
volume = {11},
journal = {Journal of Research in Personality},
doi = {10.1016/0092-6566(77)90037-X}
}

@article{Park_2020,
   title={K-EmoCon, a multimodal sensor dataset for continuous emotion recognition in naturalistic conversations},
   volume={7},
   ISSN={2052-4463},
   url={http://dx.doi.org/10.1038/s41597-020-00630-y},
   DOI={10.1038/s41597-020-00630-y},
   number={1},
   journal={Scientific Data},
   publisher={Springer Science and Business Media LLC},
   author={Park, Cheul Young and Cha, Narae and Kang, Soowon and Kim, Auk and Khandoker, Ahsan Habib and Hadjileontiadis, Leontios and Oh, Alice and Jeong, Yong and Lee, Uichin},
   year={2020},
   month=sep }

@article{SOLEYMANI20173,
title = {A survey of multimodal sentiment analysis},
journal = {Image and Vision Computing},
volume = {65},
pages = {3-14},
year = {2017},
note = {Multimodal Sentiment Analysis and Mining in the Wild Image and Vision Computing},
issn = {0262-8856},
doi = {https://doi.org/10.1016/j.imavis.2017.08.003},
url = {https://www.sciencedirect.com/science/article/pii/S0262885617301191},
author = {Mohammad Soleymani and David Garcia and Brendan Jou and Björn Schuller and Shih-Fu Chang and Maja Pantic}
}

@article{10.1145/3129340,
author = {Schuller, Bj\"{o}rn W.},
title = {Speech emotion recognition: two decades in a nutshell, benchmarks, and ongoing trends},
year = {2018},
issue_date = {May 2018},
publisher = {Association for Computing Machinery},
address = {New York, NY, USA},
volume = {61},
number = {5},
issn = {0001-0782},
url = {https://doi.org/10.1145/3129340},
doi = {10.1145/3129340},
journal = {Commun. ACM},
month = apr,
pages = {90–99},
numpages = {10}
}

@incollection{10.1093/acprof:oso/9780190613501.003.0018,
    author = {Aviezer, Hillel and Hassin, Ran},
    isbn = {9780190613501},
    title = {333Inherently Ambiguous: An Argument for Contextualized Emotion Perception},
    booktitle = {The Science of Facial Expression},
    publisher = {Oxford University Press},
    year = {2017},
    month = {04},
    doi = {10.1093/acprof:oso/9780190613501.003.0018},
    url = {https://doi.org/10.1093/acprof:oso/9780190613501.003.0018},
    eprint = {https://academic.oup.com/book/0/chapter/151285381/chapter-ag-pdf/44977033/book\_6979\_section\_151285381.ag.pdf},
}

@inproceedings{snow-etal-2008-cheap,
    title = "Cheap and Fast {--} But is it Good? Evaluating Non-Expert Annotations for Natural Language Tasks",
    author = "Snow, Rion  and
      O{'}Connor, Brendan  and
      Jurafsky, Daniel  and
      Ng, Andrew",
    editor = "Lapata, Mirella  and
      Ng, Hwee Tou",
    booktitle = "Proceedings of the 2008 Conference on Empirical Methods in Natural Language Processing",
    month = oct,
    year = "2008",
    address = "Honolulu, Hawaii",
    publisher = "Association for Computational Linguistics",
    url = "https://aclanthology.org/D08-1027/",
    pages = "254--263"
}

@article{Saganowski2022EmognitionDE,
  title={Emognition dataset: emotion recognition with self-reports, facial expressions, and physiology using wearables},
  author={Stanisław Saganowski and Joanna Komoszynska and Maciej Behnke and Bartosz Perz and Dominika Kunc and Bartłomiej Klich and Łukasz D. Kaczmarek and Przemyslaw Kazienko},
  journal={Scientific Data},
  year={2022},
  volume={9},
  url={https://api.semanticscholar.org/CorpusID:248005383}
}

@inproceedings{10.1145/2993148.2993173,
author = {Zhang, Biqiao and Essl, Georg and Mower Provost, Emily},
title = {Automatic recognition of self-reported and perceived emotion: does joint modeling help?},
year = {2016},
isbn = {9781450345569},
publisher = {Association for Computing Machinery},
address = {New York, NY, USA},
url = {https://doi.org/10.1145/2993148.2993173},
doi = {10.1145/2993148.2993173},
booktitle = {Proceedings of the 18th ACM International Conference on Multimodal Interaction},
pages = {217–224},
numpages = {8},
location = {Tokyo, Japan},
series = {ICMI '16}
}

@misc{shou2023comprehensivesurveymultimodalconversational,
      title={A Comprehensive Survey on Multi-modal Conversational Emotion Recognition with Deep Learning}, 
      author={Yuntao Shou and Tao Meng and Wei Ai and Nan Yin and Keqin Li},
      year={2023},
      eprint={2312.05735},
      archivePrefix={arXiv},
      primaryClass={cs.AI},
      url={https://arxiv.org/abs/2312.05735}, 
}

@article{ambiguitySurveyRai,
author = {Rai, Alpana and Agrawal, Chetan and Envey, Divya and Journal, Ijeasm},
year = {2025},
month = {01},
pages = {2582-6948},
title = {A Comprehensive Survey of Multimodal Emotion Recognition: Techniques, Applications, and Future Directions},
volume = {6}
}

@Article{mti6060047,
AUTHOR = {Siddiqui, Mohammad Faridul Haque and Dhakal, Parashar and Yang, Xiaoli and Javaid, Ahmad Y.},
TITLE = {A Survey on Databases for Multimodal Emotion Recognition and an Introduction to the VIRI (Visible and InfraRed Image) Database},
JOURNAL = {Multimodal Technologies and Interaction},
VOLUME = {6},
YEAR = {2022},
NUMBER = {6},
ARTICLE-NUMBER = {47},
URL = {https://www.mdpi.com/2414-4088/6/6/47},
ISSN = {2414-4088},
DOI = {10.3390/mti6060047}
}

@article{PAN2023126866,
title = {A review of multimodal emotion recognition from datasets, preprocessing, features, and fusion methods},
journal = {Neurocomputing},
volume = {561},
pages = {126866},
year = {2023},
issn = {0925-2312},
doi = {https://doi.org/10.1016/j.neucom.2023.126866},
url = {https://www.sciencedirect.com/science/article/pii/S092523122300989X},
author = {Bei Pan and Kaoru Hirota and Zhiyang Jia and Yaping Dai}
}

@article{AmbiguitySurveyPriya,
author = {Manju Priya Arthanarisamy Ramaswamy, Suja Palaniswamy},
year = {2024},
month = {08},
title = {Multimodal emotion recognition: A comprehensive review, trends, and challenges},
}

@incollection{PLUTCHIK19803,
title = {Chapter 1 - A GENERAL PSYCHOEVOLUTIONARY THEORY OF EMOTION},
editor = {Robert Plutchik and Henry Kellerman},
booktitle = {Theories of Emotion},
publisher = {Academic Press},
pages = {3-33},
year = {1980},
isbn = {978-0-12-558701-3},
doi = {https://doi.org/10.1016/B978-0-12-558701-3.50007-7},
url = {https://www.sciencedirect.com/science/article/pii/B9780125587013500077},
author = {ROBERT PLUTCHIK}
}

@INPROCEEDINGS{5349500,
  author={Mower, Emily and Metallinou, Angeliki and Lee, Chi-Chun and Kazemzadeh, Abe and Busso, Carlos and Lee, Sungbok and Narayanan, Shrikanth},
  booktitle={2009 3rd International Conference on Affective Computing and Intelligent Interaction and Workshops}, 
  title={Interpreting ambiguous emotional expressions}, 
  year={2009},
  volume={},
  number={},
  pages={1-8},
  doi={10.1109/ACII.2009.5349500}}

@misc{LikertScale,
    title = {What Is the Likert Scale? Definition, Examples, and Uses},
    howpublished = {\url{https://www.explorepsychology.com/likert-scale-definition-examples-and-uses/}},
    key = 1,
    accessed = {28/02/2025},
    year = {2025},
    author = {Explore Psychology}
}

@article{plisiecki2024highriskpoliticalbias,
      title={High Risk of Political Bias in Black Box Emotion Inference Models}, 
      author={Hubert Plisiecki and Paweł Lenartowicz and Maria Flakus and Artur Pokropek},
      year={2024},
      eprint={2407.13891},
      archivePrefix={arXiv},
      primaryClass={cs.CL},
      url={https://arxiv.org/abs/2407.13891}, 
}

@article{BussoNaturalisticDataset,
author = {Mariooryad, Soroosh and Lotfian, R. and Busso, Carlos},
year = {2014},
month = {01},
pages = {238-242},
title = {Building a naturalistic emotional speech corpus by retrieving expressive behaviors from existing speech corpora},
journal = {Proceedings of the Annual Conference of the International Speech Communication Association, INTERSPEECH}
}

@InProceedings{10.1007/978-3-642-23163-6_27,
author="Schipor, Ovidiu A.
and Pentiuc, Stefan G.
and Schipor, Maria D.",
editor="Dicheva, Darina
and Markov, Zdravko
and Stefanova, Eliza",
title="Using a Fuzzy Emotion Model in Computer Assisted Speech Therapy",
booktitle="Third International Conference on Software, Services and Semantic Technologies S3T 2011",
year="2011",
publisher="Springer Berlin Heidelberg",
address="Berlin, Heidelberg",
pages="189--193",
isbn="978-3-642-23163-6"
}

@inproceedings{Parrott2001EmotionsIS,
  title={Emotions in social psychology : essential readings},
  author={W. Gerrod Parrott},
  year={2001},
  url={https://api.semanticscholar.org/CorpusID:141721437}
}

@INPROCEEDINGS{7344624,
  author={Kim, Yelin and Provost, Emily Mower},
  booktitle={2015 International Conference on Affective Computing and Intelligent Interaction (ACII)}, 
  title={Leveraging inter-rater agreement for audio-visual emotion recognition}, 
  year={2015},
  volume={},
  number={},
  pages={553-559},
  doi={10.1109/ACII.2015.7344624}}

@InProceedings{Deng_2021_ICCV,
    author    = {Deng, Didan and Wu, Liang and Shi, Bertram E.},
    title     = {Iterative Distillation for Better Uncertainty Estimates in Multitask Emotion Recognition},
    booktitle = {Proceedings of the IEEE/CVF International Conference on Computer Vision (ICCV) Workshops},
    month     = 10,
    year      = {2021},
    pages     = {3557-3566}
}

@INPROCEEDINGS{8682170,
  author={Chou, Huang-Cheng and Lee, Chi-Chun},
  booktitle={ICASSP 2019 - 2019 IEEE International Conference on Acoustics, Speech and Signal Processing (ICASSP)}, 
  title={Every Rating Matters: Joint Learning of Subjective Labels and Individual Annotators for Speech Emotion Classification}, 
  year={2019},
  volume={},
  number={},
  pages={5886-5890},
  doi={10.1109/ICASSP.2019.8682170}}

@misc{hartmann2022emotionenglish,
  author={Hartmann, Jochen},
  title={Emotion English DistilRoBERTa-base},
  year={2022},
  howpublished = {\url{https://huggingface.co/j-hartmann/emotion-english-distilroberta-base/}},
}

@article{Scherer1994,
  title = {Evidence for universality and cultural variation of differential emotion response patterning.},
  volume = {66},
  ISSN = {0022-3514},
  url = {http://dx.doi.org/10.1037/0022-3514.66.2.310},
  DOI = {10.1037/0022-3514.66.2.310},
  number = {2},
  journal = {Journal of Personality and Social Psychology},
  publisher = {American Psychological Association (APA)},
  author = {Scherer,  Klaus R. and Wallbott,  Harald G.},
  year = {1994},
  pages = {310–328}
}

@misc{khoa2024emotionenglishAudio,
  author={Khoa},
  title={Wav2Vec2 Speech Emotion Recognition for English},
  year={2024},
  howpublished = {\url{https://huggingface.co/Khoa/w2v-speech-emotion-recognition}},
}

@article{RYUMINA2022,
  title        = {In Search of a Robust Facial Expressions Recognition Model: A Large-Scale Visual Cross-Corpus Study},
  author       = {Elena Ryumina and Denis Dresvyanskiy and Alexey Karpov},
  journal      = {Neurocomputing},
  year         = {2022},
  doi          = {10.1016/j.neucom.2022.10.013},
  url          = {https://www.sciencedirect.com/science/article/pii/S0925231222012656},
}

@misc{hu2024recenttrendsmultimodalaffective,
      title={Recent Trends of Multimodal Affective Computing: A Survey from NLP Perspective}, 
      author={Guimin Hu and Yi Xin and Weimin Lyu and Haojian Huang and Chang Sun and Zhihong Zhu and Lin Gui and Ruichu Cai and Erik Cambria and Hasti Seifi},
      year={2024},
      eprint={2409.07388},
      archivePrefix={arXiv},
      primaryClass={cs.CL},
      url={https://arxiv.org/abs/2409.07388}, 
}

@book{FERBook,
author = {Verma, Gyanendra},
year = {2023},
month = {03},
pages = {},
title = {Multimodal Affective Computing: Affective Information Representation, Modelling, and Analysis},
isbn = {9789815124453},
doi = {10.2174/97898151244531230101}
}

@misc{ma2023emotion2vecselfsupervisedpretrainingspeech,
      title={emotion2vec: Self-Supervised Pre-Training for Speech Emotion Representation}, 
      author={Ziyang Ma and Zhisheng Zheng and Jiaxin Ye and Jinchao Li and Zhifu Gao and Shiliang Zhang and Xie Chen},
      year={2023},
      eprint={2312.15185},
      archivePrefix={arXiv},
      primaryClass={cs.CL},
      url={https://arxiv.org/abs/2312.15185}, 
}

@inproceedings{Preeti2012MULTIMODALER,
  title={MULTIMODAL EMOTION RECOGNITION FOR ENHANCING HUMAN COMPUTER INTERACTION},
  author={Khanna Preeti},
  year={2012},
  url={https://api.semanticscholar.org/CorpusID:260521063}
}

@article{MultimodalEmotionRecognitionusingDeepLearning_2021, volume={2}, url={https://jastt.org/index.php/jasttpath/article/view/91}, DOI={10.38094/jastt20291}, abstractNote={
New research into human-computer interaction seeks to consider the consumer’s emotional status to provide a seamless human-computer interface. This would make it possible for people to survive and be used in widespread fields, including education and medicine. Multiple techniques can be defined through human feelings, including expressions, facial images, physiological signs, and neuroimaging strategies. This paper presents a review of emotional recognition of multimodal signals using deep learning and comparing their applications based on current studies. Multimodal affective computing systems are studied alongside unimodal solutions as they offer higher accuracy of classification. Accuracy varies according to the number of emotions observed, features extracted, classification system and database consistency. Numerous theories on the methodology of emotional detection and recent emotional science address the following topics. This would encourage studies to understand better physiological signals of the current state of the science and its emotional awareness problems.
}, number={01}, journal={Journal of Applied Science and Technology Trends}, year={2021}, month={May}, pages={73–79}, author={Sharmeen M.Saleem Abdullah Abdullah}}

@article{SystematicAdressing,
author = {García-Hernández, Rosa and Luna-Garcia, Huizilopoztli and Celaya Padilla, Jose and García, Alejandra and Reveles, Luis and Flores-Chaires, Luis and Delgado Contreras, Juan Ruben and Rondon, David and Villalba, Klinge},
year = {2024},
month = {08},
pages = {7165},
title = {A Systematic Literature Review of Modalities, Trends, and Limitations in Emotion Recognition, Affective Computing, and Sentiment Analysis},
volume = {14},
journal = {Applied Sciences},
doi = {10.3390/app14167165}
}

@inproceedings{MELD,
  title={MELD: A Multimodal Multi-Party Dataset for Emotion Recognition in Conversations},
  author={Poria, Soujanya and Hazarika, Devamanyu and Majumder, Navonil and Naik, Gautam and Cambria, Erik and Mihalcea, Rada},
  booktitle={Proceedings of the ACL 2019},
  year={2019}
}

@article{EmoDB,
  author = {Burkhardt, Felix and Paeschke, Angelika and Rolfes, Michael and Sendlmeier, Werner F. and Weiss, Benjamin},
  title = {A Database of German Emotional Speech},
  journal = {Interspeech},
  year = {2005}
}

@misc{TESS,
  title = {Toronto emotional speech set (TESS)},
  author = {Dupuis, Kate and Pichora-Fuller, M. Kathleen},
  year = {2010},
  note = {Available: \url{https://tspace.library.utoronto.ca/handle/1807/24487}}
}

@inproceedings{SAVEE,
  author = {Haq, S. and Jackson, P. J. B.},
  title = {Audio-visual emotion recognition using AdaBoost},
  booktitle = {Proceedings of the International Conference on Auditory-Visual Speech Processing},
  year = {2009}
}

@inproceedings{eNTERFACE,
  title={eNTERFACE'05 Audio-Visual Emotion Database},
  author={Martin, Olivier and Kotsia, Irene and Macq, Benoit and Pitas, Ioannis},
  booktitle={Data Challenge Workshop on Emotion Recognition},
  year={2005}
}

@article{FER2013,
  author = {Goodfellow, Ian J. and Erhan, Dumitru and Carrier, Pierre and Courville, Aaron C. and Mirza, Mehdi and Hamner, Ben and Cukierski, William and Tang, Yuan and Thaler, David and Lee, Dong-Hyun and Zhou, Ying and Ramaiah, Chuan and Feng, Yan and Li, Rui and Wang, Xiaoguang and Sherstinsky, Alex and Tang, Tony},
  title = {Challenges in Representation Learning: A Report on Three Machine Learning Contests},
  journal = {Neural Information Processing},
  year = {2013},
  note = {FER-2013 described in the ICML 2013 Challenges in Representation Learning},
}

@Article{app12020807,
AUTHOR = {Xiao, Huafei and Li, Wenbo and Zeng, Guanzhong and Wu, Yingzhang and Xue, Jiyong and Zhang, Juncheng and Li, Chengmou and Guo, Gang},
TITLE = {On-Road Driver Emotion Recognition Using Facial Expression},
JOURNAL = {Applied Sciences},
VOLUME = {12},
YEAR = {2022},
NUMBER = {2},
ARTICLE-NUMBER = {807},
URL = {https://www.mdpi.com/2076-3417/12/2/807},
ISSN = {2076-3417},
DOI = {10.3390/app12020807}
}

@article{DBLP:journals/corr/abs-2110-14957,
  author       = {Th{\'{e}}o Deschamps{-}Berger and
                  Lori Lamel and
                  Laurence Devillers},
  title        = {End-to-End Speech Emotion Recognition: Challenges of Real-Life Emergency
                  Call Centers Data Recordings},
  journal      = {CoRR},
  volume       = {abs/2110.14957},
  year         = {2021},
  url          = {https://arxiv.org/abs/2110.14957},
  eprinttype    = {arXiv},
  eprint       = {2110.14957},
  bibsource    = {dblp computer science bibliography, https://dblp.org}
}

@article{EspinoSalinas2024,
  title = {Multimodal driver emotion recognition using motor activity and facial expressions},
  volume = {7},
  ISSN = {2624-8212},
  url = {http://dx.doi.org/10.3389/frai.2024.1467051},
  DOI = {10.3389/frai.2024.1467051},
  journal = {Frontiers in Artificial Intelligence},
  publisher = {Frontiers Media SA},
  author = {Espino-Salinas,  Carlos H. and Luna-García,  Huizilopoztli and Celaya-Padilla,  José M. and Barría-Huidobro,  Cristian and Gamboa Rosales,  Nadia Karina and Rondon,  David and Villalba-Condori,  Klinge Orlando},
  year = {2024},
  month = nov 
}

@misc{PeterNeville,
  author       = {Peter Neville},
  title        = {Personal Communication},
  howpublished = {Interview},
  year         = {2025},
  note         = {Interview held february 2025}
}

@INPROCEEDINGS{9746450,
  author={Das, Sneha and Nadine Lønfeldt, Nicole and Katrine Pagsberg, Anne and Clemmensen, Line H.},
  booktitle={ICASSP 2022 - 2022 IEEE International Conference on Acoustics, Speech and Signal Processing (ICASSP)}, 
  title={Towards Transferable Speech Emotion Representation: On Loss Functions for Cross-Lingual Latent Representations}, 
  year={2022},
  volume={},
  number={},
  pages={6452-6456},
  doi={10.1109/ICASSP43922.2022.9746450}}

@inproceedings{das2022continuous,
  title={Continuous Metric Learning For Transferable Speech Emotion Recognition and Embedding Across Low-resource Languages},
  author={Das, Sneha and Lund, Nicklas Leander and L{\o}nfeldt, Nicole Nadine and Pagsberg, Anne Katrine and Clemmensen, Line Katrine Harder},
  booktitle={Northern Lights Deep Learning Workshop 2022},
  year={2022}
}

@inproceedings{das2022zero,
  title={Zero-shot Cross-lingual Speech Emotion Recognition: A Study of Loss Functions and Feature Importance},
  author={Das, Sneha and Lonfeldt, Nicole Nadine and Lund, Nicklas Leander and Pagsberg, Anne Katrine and Clemmensen, Line Katrine Harder},
  booktitle={2nd Symposium on Security and Privacy in Speech Communication},
  year={2022}
}

@inproceedings{hjuler2025exploring,
  title={Exploring local interpretable model-agnostic explanations for speech emotion recognition with distribution-shift},
  author={Hjuler, Maja J and Clemmensen, Line H and Das, Sneha},
  booktitle={ICASSP 2025-2025 IEEE International Conference on Acoustics, Speech and Signal Processing (ICASSP)},
  pages={1--5},
  year={2025},
  organization={IEEE}
}

@article{hjuler2025emotale,
  title={Emotale: An enacted speech-emotion dataset in danish},
  author={Hjuler, Maja J and Skat-R{\o}rdam, Harald V and Clemmensen, Line H and Das, Sneha},
  journal={arXiv preprint arXiv:2508.14548},
  year={2025}
}
\section{Appendix}
\begin{figure}[!ht]
\centering
\begin{multicols}{2}
\begin{itemize}
    \item Affective Dissonance
    \item Affective Misalignment
    \item Affective Polarity Conflict
    \item Ambivalent Sentiment Analysis
    \item Annotator Idiosyncrasy
    \item Cognitive Dissonance
    \item Conflicting Sentiment Cues
    \item Contextual Emotion Variability
    \item Emotion Contradiction
    \item Emotion Multimodality Challenges
    \item Emotion Recognition Uncertainty
    \item Emotional Ambiguity
    \item Emotional Ambivalence
    \item Emotional Incongruity
    \item Expressive Modality Conflict
    \item Implicit vs. Explicit Emotion Mismatch
    \item Irony Detection
    \item Modality-Specific Discrepancy
    \item Multimodal Emotion Discrepancy
    \item Multimodal Sentiment Divergence
    \item Prototypicality / Non-prototypicality
    \item Sarcasm Detection
    \item Sentiment Incongruence
    \item Sentimental Misalignment
    \item Subjective Emotion Divergence
    \item Subtle Emotional Expression Analysis
    \item Uncertain Sentiment Prediction
\end{itemize}
\end{multicols}
\caption{\textbf{Terminology Related to Emotional Ambiguity and Sentiment Divergence.} \\
A non-exhaustive list of terms encountered or explored in the context of affective computing, emotion recognition, and multimodal sentiment analysis. All terms are used in their respective mentions to describe what could be placed under interprative uncertainty / ambiguity. These terms reflect the conceptual diversity and lack of consensus in labeling ambiguous or conflicting emotional expressions across modalities and annotation perspectives.}
\label{fig:emotion_ambiguity_terms}
\end{figure}

\begin{figure}[!ht]
\centering
\noindent\fbox{%
\begin{minipage}{1\textwidth}
\begin{enumerate}
    \item Which emotional categories are most commonly used to describe psychological states in clinical settings?
    \item How are emotions such as disgust or surprise, which are less commonly emphasized in clinical psychology, typically interpreted or treated across different disciplines?
    \item From a psychological standpoint, what is the typical distribution of expressed emotions in therapeutic or AI coaching contexts? Are individuals more likely to express neutrality, positivity, or negativity?
    \item To what extent is emotional ambivalence considered a natural or expected phenomenon in clinical interactions?
    \item How should the presence of conflicting emotional cues across different modalities be interpreted diagnostically?
    \item In clinical practice, how are different modalities prioritized when assessing a patient’s emotional state? Is verbal content weighed more heavily than vocal tone or facial expression?
    \item Are there expected patterns in how emotional states evolve? For example, do transitions typically move through phases such as happy → neutral → sad?
    \item If facial expressions pass through a neutral phase while shifting from one emotion to another, should this intermediate state be interpreted as genuine neutrality?
    \item What is the general clinical view on emotional neutrality? Is it often a meaningful state or more of a fallback category in ambiguous scenarios?
\end{enumerate}
\end{minipage}%
}
\caption{\textbf{Interview Protocol for Clinical Insight.} \\
List of questions posed to a clinical psychologist, and researchers in the field, to guide interpretation of emotional expression and modality prioritization in affective computing research.}
\label{fig:interview_questions}
\end{figure}

\begin{figure}[!ht]
    \centering
    \includegraphics[width=0.95\linewidth]{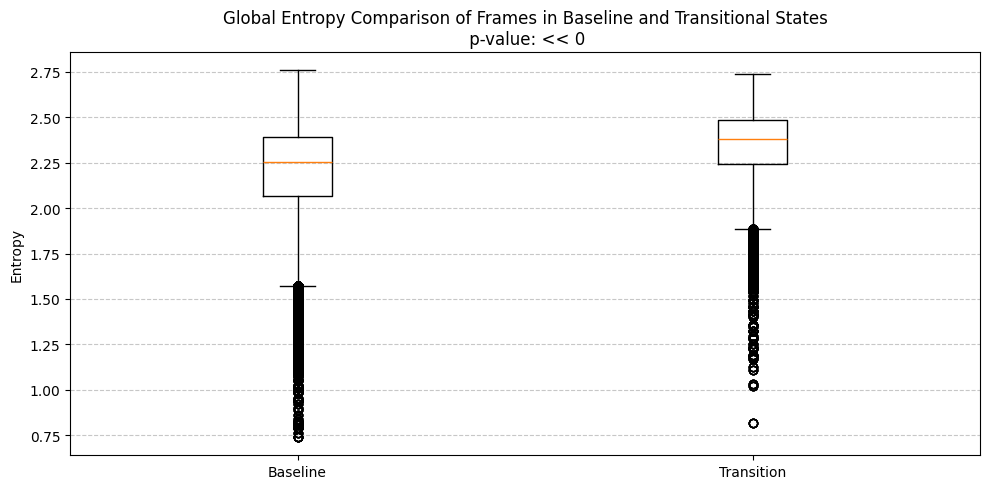}
    \caption{
    Comparison of entropy distributions across baseline and transitional frames. 
    Transitional segments—defined as regions surrounding changes in dominant emotion—exhibit significantly higher entropy than stable segments, indicating greater ambiguity during emotional shifts. Statistical testing confirms a robust difference (p $<<$ 0), supporting the claim that transitions are periods of increased affective uncertainty.
    }
    \label{fig:entropy_global}
\end{figure}

\begin{figure}[!ht]
    \centering
    \includegraphics[width=0.95\linewidth]{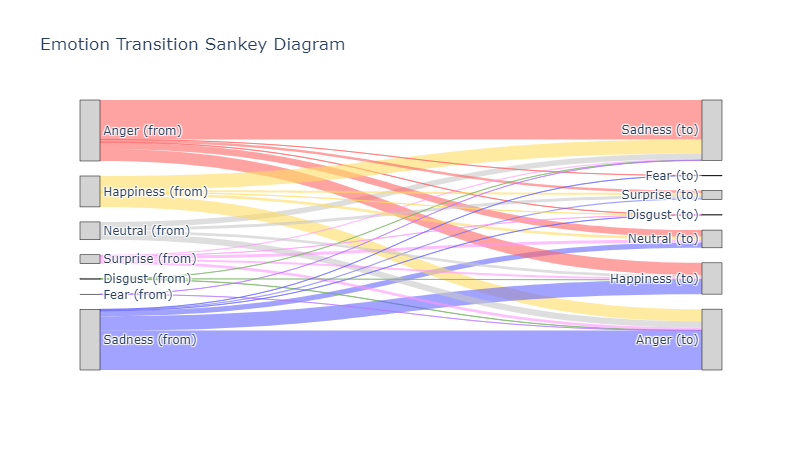}
    \caption{
    Sankey diagram showing directional frequency of emotion transitions across utterances.
    The width of each flow represents how often a given emotion transitions into another (e.g., \textit{happiness}~$\rightarrow$~\textit{surprise}, \textit{sadness}~$\rightarrow$~\textit{anger}). The structure and asymmetry of these flows shows how there are no emotions directly acting as transition emotions in the facial expressions picked up by the model.
    }
    \label{fig:sankey_diagram}
\end{figure}

\begin{figure}[!ht]
\centering
\begin{minipage}[t]{0.48\textwidth}
    \centering
    \includegraphics[width=\linewidth]{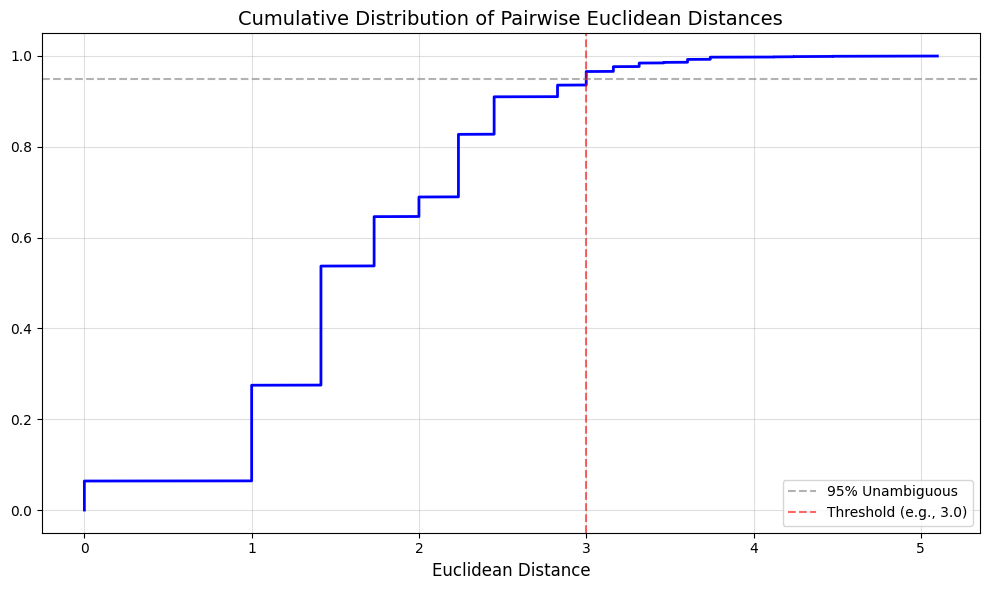}
\end{minipage}
\hfill
\begin{minipage}[t]{0.48\textwidth}
    \centering
    \includegraphics[width=\linewidth]{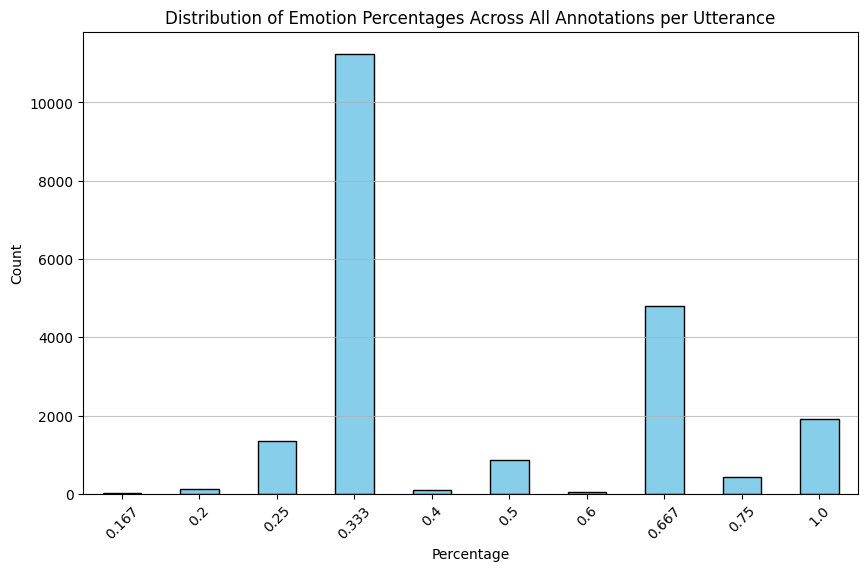}
\end{minipage}
\caption{
Illustrative distributions underlying threshold selections for ambiguity filtering. 
\textbf{Left}: Cumulative distribution of pairwise Euclidean distances across VAD annotations, used to define VAD-based agreement thresholds. The red dashed line indicates a potential cutoff (e.g., distance = 3.0) that captures 95\%. 
\textbf{Right}: Distribution of categorical agreement percentages across utterances. Most utterances fall near high disagreement (0.333), moderate agreement (0.667), or full agreement (1.0), supporting the partitioning schema used in CEA-based filtering.
}
\label{fig:thresholding_distributions}
\end{figure}

\begin{table}[!ht]
\centering
\caption{Weighted F1 scores for each modality across categorical emotion annotation (CEA) agreement thresholds. Higher thresholds reflect stricter filtering (greater annotator consensus).}
\label{tab:cea_f1_scores}
\begin{tabular}{lcccc}
\toprule
\textbf{Modality} & \textbf{No Filtering (0)} & \textbf{0.6} & \textbf{0.7} & \textbf{Strictest (1.0)} \\
\midrule
Text   & 0.366 & 0.373 & 0.376 & 0.394 \\
Audio  & 0.332 & 0.332 & 0.496 & 0.480 \\
Facial & 0.395 & 0.396 & 0.508 & 0.479 \\
\bottomrule
\end{tabular}
\end{table}

\begin{table}[!ht]
\centering
\caption{Weighted F1 scores for each modality across valence-arousal-dominance (VAD) agreement filtering levels. Higher steps correspond to stricter filtering based on annotator VAD coherence.}
\label{tab:vad_f1_scores}
\begin{tabular}{lcccccc}
\toprule
\textbf{Modality} & \textbf{No Filtering (5)} & \textbf{4} & \textbf{3} & \textbf{2} & \textbf{1} & \textbf{Strictest (0)} \\
\midrule
Text   & 0.366 & 0.366 & 0.365 & 0.384 & 0.391 & 0.378 \\
Audio  & 0.332 & 0.332 & 0.329 & 0.327 & 0.311 & 0.289 \\
Facial & 0.395 & 0.396 & 0.395 & 0.401 & 0.383 & 0.419 \\
\bottomrule
\end{tabular}
\end{table}

\begin{figure}[!ht]
    \centering
    \includegraphics[width=1.0\linewidth]{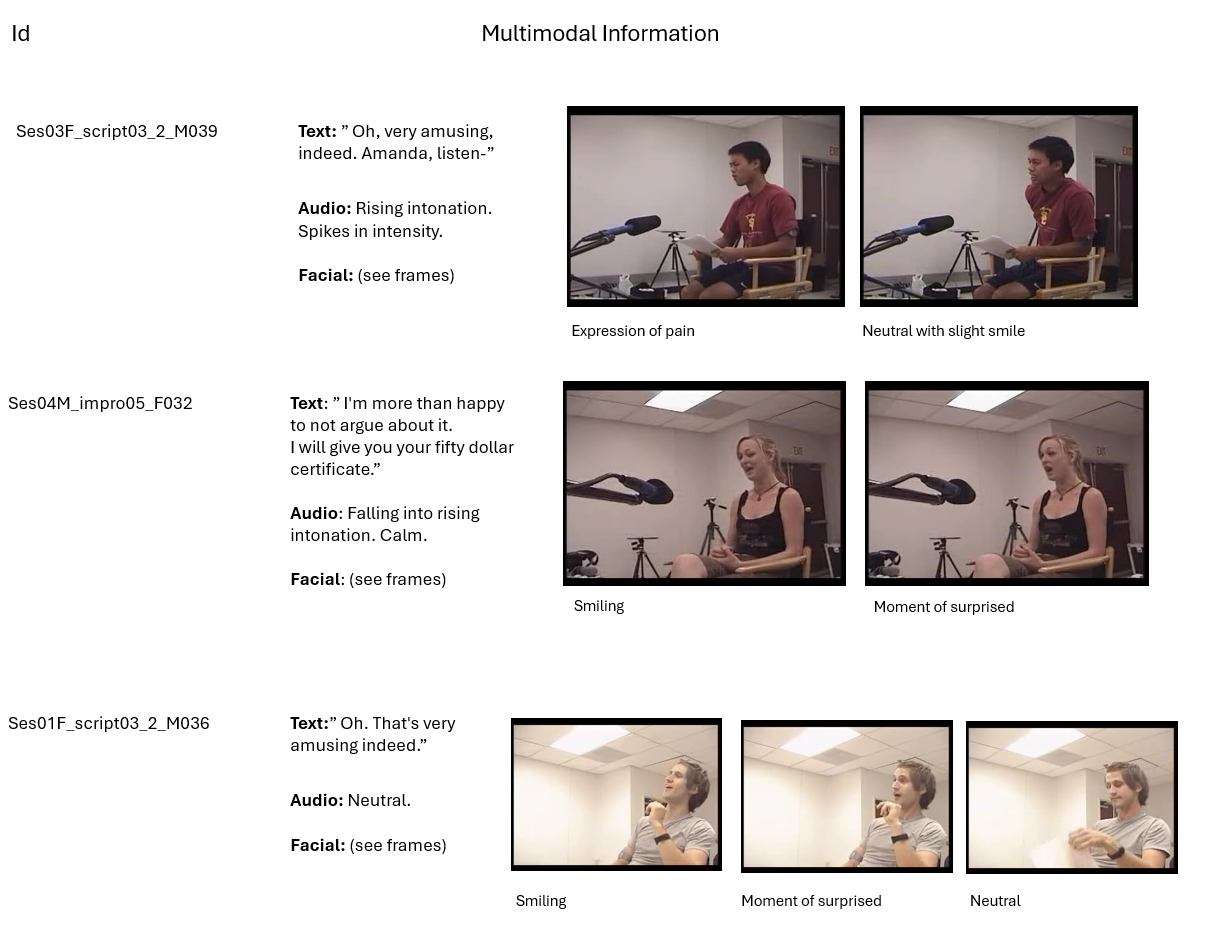}
    \caption{Extra cases of modality agreement in system predictions diverging from ground truth. All models predicted \textit{happy}, while annotators labeled the utterance as \textit{angry} or \textit{frustrated}. Examples directly taken from IEMOCAP data set \citep{Busso2008} and used under their licensing agreement.}
    \label{fig:modality-agree-fails}
\end{figure}

\end{document}